\documentclass[sigconf]{acmart}
\copyrightyear{2021}
\acmYear{2021}
\acmConference{}%[CIKM '21]{30th ACM International Conference on Information and Knowledge Management}{November 1--5, 2021}{Queensland, Australia}
\acmBooktitle{}%30th ACM International Conference on Information and Knowledge Management (CIKM '21), November 1--5, 2021, Queensland, Australia}
\acmDOI{}
\acmISBN{}

\AtBeginDocument{%
  \providecommand\BibTeX{{%
    \normalfont B\kern-0.5em{\scshape i\kern-0.25em b}\kern-0.8em\TeX}}}

\usepackage{siunitx}

\usepackage{float}
\usepackage{soul}
\usepackage{color}
\usepackage{amsmath}
\usepackage{bbm}
\usepackage{algorithm}
\usepackage{algpseudocode}

\newcommand{\loss}{\mathcal{L}}

\definecolor{Blue}{rgb}{0,0,1}

\definecolor{Orange}{rgb}{1,0.5,0}

\definecolor{Green}{rgb}{0,1,0}

\DeclareMathOperator*{\argmin}{argmin}

\begin{document}

\sisetup{tight-spacing=true}

\author{Javier Albert}
\email{javier.albert@booking.com}
\affiliation{\institution{Booking.com, Tel Aviv, Israel}}
\authornote{Both authors contributed equally to this research.}

\author{Dmitri Goldenberg}
\email{dima.goldenberg@booking.com}
\affiliation{\institution{Booking.com, Tel Aviv, Israel}}
\authornotemark[1]

%% revising paper checklist:
%% https://web.cs.dal.ca/~eem/gradResources/KDD/Checklist%20for%20Revising%20a%20SIGKDD%20Data%20Mining%20Paper.pdf

\begin{abstract}

Promotions and discounts are essential components of modern e-commerce platforms, where they are often used to incentivize customers towards purchase completion. Promotions also affect revenue and may incur a monetary loss that is often limited by a dedicated promotional budget. We study the \emph{Online Constrained Multiple-Choice Promotions Personalization Problem}, where the optimization goal is to select for each customer which promotion to present in order to maximize purchase completions, while also complying with global budget limitations.
Our work formalizes the problem as an Online Multiple Choice Knapsack Problem and extends the existent literature by addressing cases with negative weights and values. We provide a real-time adaptive method that guarantees budget constraints compliance and achieves above 99.7\% of the optimal promotional impact on various datasets. Our method is evaluated on a large-scale experimental study at Booking.com - one of the leading online travel platforms in the world.

\end{abstract}





\title{E-Commerce Promotions Personalization via Online Multiple-Choice Knapsack with Uplift Modeling}

\maketitle

\section{Introduction}\label{sec:intro}

Promotional offers, such as discounts and rewards, play a key role in marketing efforts of online e-commerce platforms. Promotions are expected to generate a significant uplift in sales, offering more value to the customers and are often used to drive customer base growth \cite{hansotia2002direct}.
While increasing the probability to purchase, a promotion can also incur an incremental monetary loss if the net revenue from sales when the promotion is offered is smaller than the net revenue from sales when the promotion is not offered \cite{lin2017monetary,zhao2019unified}. A dedicated budget usually limits this incremental net revenue loss. The promotional campaign can remain sustainable as long as the overall incremental net revenue loss is within this budget \cite{goldenberg2020free}.

The effect a promotion has on the probability of completing a purchase and on the expected net revenue loss varies from customer to customer. 
The \textit{Conditional Average Treatment Effect} (CATE) is defined as the expected change in a metric of interest (conversion, revenue, click rate, churn, etc.) \textit{caused} by a treatment, given the individual's characteristics \cite{devriendt2018literature}. Various machine learning techniques were introduced to estimate the CATE  \cite{athey2015machine}, and the field is commonly known as Uplift Modeling.
The input data is usually obtained from randomized controlled trials \cite{kaufman2017democratizing}, where the customers are randomly assigned with different treatments. The collected data is used to train machine learning models that estimate the incremental causal effects of each treatment at the individual level~\cite{holland1986statistics}.

A notable modeling technique for CATE estimation is the Meta-learning group of estimators, such as the two-models estimator \cite{hansotia2002direct}, the transformed outcome estimator \cite{athey2015machine} and the X-learner estimator \cite{kunzel2019metalearners}. They allow using classical machine learning techniques to estimate CATE. Other tailored methods, such as uplift trees \cite{rzepakowski2012decision} and neural network based approaches \cite{johansson2016learning,louizos2017causal,yoon2018ganite} modify existing machine learning algorithms to be suitable for CATE estimation. 
Over the recent years, uplift modeling has become popular in web and e-commerce applications, such as at Facebook, Amazon \cite{makhijani2019lore}  Criteo \cite{diemert2018large}, Uber \cite{zhao2019uplift} and Booking.com \cite{teinemaa2021uplift}.
Such models can be used for personalization purposes \cite{goldenberg2021personalization} since we can use the estimations to decide if a customer should be treated or not. They are also used to target a specific segment of the customer base with a promotional offer or other types of marketing campaigns. Personalization solutions in online marketplaces often address multi-objective problems \cite{mehrotra2020advances}. Given the budget constraints typical to a promotional offering, we are usually interested in targeting the promotion to maximize the total incremental effect on sales while we also comply with the overall budget constraint. This optimization problem takes the form of a 0-1 knapsack problem \cite{toth1990knapsack}.

In previous work \cite{goldenberg2020free} we addressed the causal estimation and binary promotion allocation problem, where we introduced the \textit{Retrospective Estimation} technique. 
The proposed solution presents a dynamic system, similar to the Online Knapsack problem \cite{lueker1998average}, that can decide whether a promotion should be assigned to a customer based on the CATE estimations. It allows dynamic calibration based on the overall measured impact without harming the experience of an individual customer \cite{goldenberg2021learning}. This framework addresses the fact that a promotional offer can result in both a positive and negative incremental net revenue loss and thus introduces a knapsack problem with negative weights.
Similar work on cost-sensitive uplift decision making was done on synthetic \cite{olaya2021or} and real \cite{du2019improve,miller2020personalized} promotional marketing use-cases. Such solutions often address a single-offer treatment, which need to account for a direct comparison between two alternatives - \textit{to treat or not to treat?} \cite{rossler2021treat}.

For this general case, the incremental comparison is not trivial anymore. Should the incremental uplift be compared to the no-promotion baseline or the second-best option?
A recent work tackled the multiple-promotions uplift problem with novel meta-learning techniques \cite{zhao2019uplift}, taking into account both conversion and cost.
Another promotions recommendation work suggested an offline constrained optimization solution \cite{makhijani2019lore}, tackling the limited marketing budget problem and assuming a constant cost per promotion.
However, such solutions are limited in a realistic e-commerce setup, where the decision about promotion selection under up-to-date budget constraints needs to be made online. One possible solution for this online challenge could be achieved using Bandits with Knapsacks \cite{tran2012knapsack, badanidiyuru2013bandits}. Such methods might limit the ability to perform incremental estimations, such as Return on Investment (ROI) measurements and do not account for incremental negative effects on value or weight.

A potential solution for a two-steps estimation and optimization approach can arise from the multiple-choice knapsack problem (MCKP), specifically, the online MCKP \cite{zhou2008budget, zhou2008algorithm}. 
Similarly to the described case, the optimization goal is to pick a single promotion at each decision point (i.e., customer visit), given both the value and weight quantities.
This online formulation allows for dynamic strategy adaptation, which is a common practice in marketing use-cases~\cite{fischer2011dynamic,sela2016spending,lin2017monetary}.

This paper analyzes the budget constrained multiple-choice promotion assignment problem, proposes an online solution, and compares it to various benchmarks in a real-life experimental study. Our main contributions are:
\begin{enumerate}
    \item Formulation the \textit{Online Constrained Multiple-Choice Promotions Personalization Problem} as an Online-MCKP.
    \item A two-steps solution, relying on uplift modeling estimations and multiple-choice constrained optimization.
    \item Extension of Online-MCKP solution with negative values and weights.
    \item A Large-scale experimental study on real promotions offering, conducted on Booking.com platform.
\end{enumerate}

The study proceeds as follows: \autoref{sec:formulation} describes and formalizes the  \textit{Online Constrained Multiple-Choice Promotions Personalization Problem}. 
\autoref{sec:solution} covers the solution framework, including the uplift modeling method, linear programming solution, and suggested Online-MCKP approach.
\autoref{sec:experiment} describes a large-scale experimental study, including details on datasets and methods, and the experimental setup.
\autoref{sec:results} presents and discusses the experimental results.
The final section concludes.

\begin{figure}[b]
    \includegraphics[width=0.95\columnwidth]{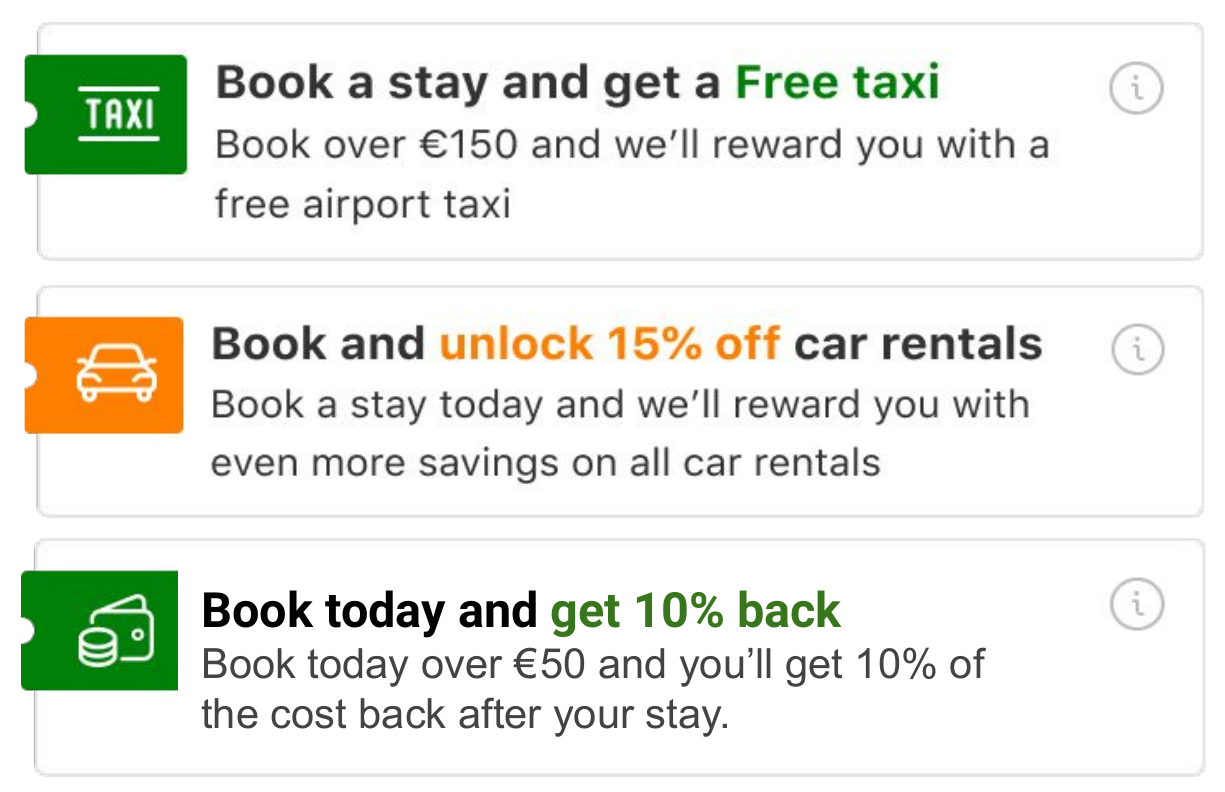}
        \caption{Examples of travel promotions.}
        \Description{Examples of travel promotions: a free taxi, discount on car rental or 10\% money return}
    \label{fig:promo_example}
    \centering
    % source: https://docs.google.com/presentation/d/13MQZgr_69YbTVsWMY4z_dKOQPoFkSVUuvir0UgAXqH4/edit#slide=id.gd600a9cd7e_0_788
\end{figure}

\section{Problem formulation}
\label{sec:formulation}

\begin{figure*}[t]
\centering
\includegraphics[width=0.99\textwidth]{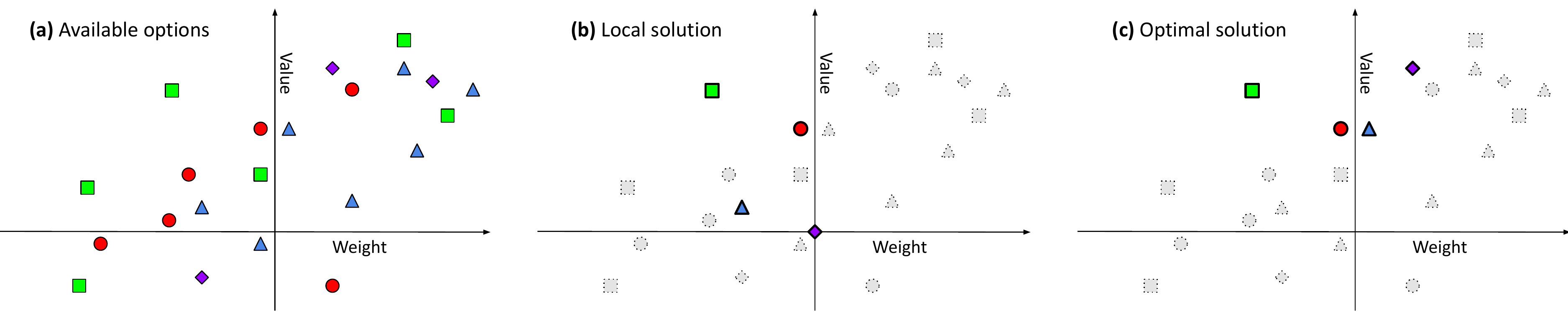}
 \caption{Toy example of promotions assignment. (a) presents all available promotion options of 4 customers (colors and shapes) on value and weight axes; (b) presents a local per-customer solution (colored); (c) presents the global optimal selection.}
 \Description{Toy example of promotion selection optimization. (a) presents all available promotion options of 4 customers (colors) on value and weight axes; (b) presents a local per-customer solution (colored); (c) presents the global optimal selection}
  \label{fig:toy_example}
%source 
%https://docs.google.com/presentation/d/16vDuAaivklhXFYAJDGucZE_nBnUC45m4E8A5azzoRIY/edit?usp=sharing
\end{figure*}

In this work, we address the personalized assignment of promotional offers on an online e-commerce platform. The optimization target is to maximize the overall incremental number of customers completing a purchase.
We are allowed to pick at most one promotion to offer each customer from a finite set of eligible promotions (see an example of potential promotions in the travel industry in \autoref{fig:promo_example}). A global budget constrains the overall incremental net revenue loss generated by the promotions.

\subsection{Treatment Effect Estimation}
For each customer $i$ in the customers set $U$ we define $Y_i$ - a binary random variable representing a completion of a purchase and $R_i$ - a continuous random variable representing the net monetary revenue associated with the purchase (sum of all revenues minus all promotional costs).
We adopt the Potential Outcomes framework \cite{imbens2010rubin}, which allows us to express the causal effects of the promotions on these two variables.  $Y_i(k)$ for a customer $i$ and a promotion $k$ represents the potential purchase if customer $i$ is offered the promotion $k$, while $R_i(k)$ represents the potential net revenue if customer $i$ is offered the promotion $k$. Likewise, $Y_i(0)$ and $R_i(0)$ represent the potential outcomes if no promotion is offered to customer $i$, an option that is always available for every customer. We can thus define the conditional average treatment effect on $Y_i$ and $R_i$ for a customer with pre-promotion covariates $x$ as follows: 

$$\mathit{CATE}_{Y}(i,k) = \mathbf{E}(Y_i(k)-Y_i(0)\mid X=x_i)$$
$$\mathit{CATE}_{R}(i,k) =\mathbf{E}(R_i(k)-R_i(0)\mid X=x_i)$$
$\mathit{CATE}_{Y}(i,k)$ represents the incremental effect on the expected purchase probability of customer $i$ if presented with promotion $k$, while $\mathit{CATE}_{R}(i,k)$ represents the incremental effect on the expected net revenue. 
We assume that both $\mathit{CATE}_{Y}(i,k)$ and $\mathit{CATE}_{R}(i,k)$ can be either positive or negative. 
The conditional treatment effect on the expected net revenue loss $\loss$ is defined as:  
$$\mathit{CATE}_{\loss}(i,k) = -\mathit{CATE}_{R}(i,k)$$
For each customer $i$ and each of the possible promotions $k \in K_i$ we can estimate $\mathit{CATE}_{Y}(i,k)$ and $\mathit{CATE}_{\loss}(i,k)$ using uplift modeling on data gathered from randomized controlled trials, similarly to our previous work \cite{goldenberg2020free}.  
The base item $k=0$ for which $\mathit{CATE}_{Y}(i,0) = 0$ and $\mathit{CATE}_{\loss}(i,0) = 0$ represents no promotion. The optimization goal is to choose for every customer $i$ a single element $k^* \in K_i$ in order to maximize the sum of all the selected $\mathit{CATE}_{Y}(i,k^*)$, while the sum of all the selected $\mathit{CATE}_{\loss}(i,k^*)$ does not exceed the budget constraint $C$.

\subsection{Multiple-Choice Knapsack Problem}

Formally, the presented problem can be described as follows:
\begin{equation}
\label{eq:mckp}
\begin{array}{ll}
\noindent
\text{\textbf{Maximize }}
\displaystyle\sum_{i\in U}\sum_{k\in K_i}
 \mathit{CATE}_{Y}(i,k) \cdot Z_{ik}  &\\
\noindent

\text{\textbf{subject to:}}&\\\\
1.\;
\displaystyle\sum_{i\in U}\sum_{k\in K_i}
 \mathit{CATE}_{\loss}(i,k) \cdot Z_{ik} \leq C  & \forall i\in U, k\in K_i \\\\

2.\;  \displaystyle\sum_{k\in K_i}  Z_{ik} = 1   & \forall i\in U \\\\

3.\:\:  Z_{ik} \in \{0,1\} & \forall i\in U, k \in K_i\\
\end{array}
\end{equation}
Here, $Z_{i,k}$ is a binary assignment variable indicating whether a customer $i$ is offered the promotion $k$ or not. The linear formulation above  matches the Multiple-Choice knapsack problem (MCKP) \cite{sinha1979multiple}, a variation of the known 0-1 knapsack problem.
In our case, the value of each item (a promotion offered to a specific customer) $v_{ik}$ is $\mathit{CATE}_{Y}(i,k)$ and the weight of each item $w_{ik}$ is $\mathit{CATE}_{\loss}(i,k)$. Different from the classical setup, we allow the weights and values of the items to be negative.
%\todo{Previous work has shown efficient algorithms \cite{dyer1984n} for solving the MCKP in an offline manner.}
For practical applications, we investigate the Online-MCKP \cite{zhou2008budget}, a variation of the MCKP where customers arrive one by one. Here, we need to decide which promotion to offer each customer on an online manner.

\subsection{Problem Toy Example}

We demonstrate the optimization problem by illustrating a toy example on \autoref{fig:toy_example}. In this example, we observe various promotional offers for four different customers (green-squares, red-circles, blue-triangles, and purple-rhombuses) given a budget of zero. In other words - we need to pick one promotion per customer, such that the total weight (expected net revenue loss $\loss$) will not be positive.
For each promotion $k$ we present its value $\mathit{CATE}_{Y}(i,k)$ and weight $\mathit{CATE}_{\loss}(i,k)$  on a two-dimensional chart, as shown in sub-figure (a). We observe promotions in all four quadrants of the axes, representing both positive and negative expected value (y-axis) and weight (x-axis).
Sub-figure (b) presents a local solution to the problem - for each customer, we pick the promotion with the highest value and non-positive weight. This solution results in picking promotions from one quadrant of the axes only - where the value is positive, and weight is negative. It is important to note that there is no purple option in the quadrant, and therefore we defaulted for the base $(0,0)$ solution - where we do not offer any promotion to the customer.
Sub-figure (c) presents the optimal solution to the problem. In this case, we also pick promotions with positive weight for the blue and purple customers since their weight is compensated with the negative weight of the selected green and red promotions. We can observe that the total value (overall position of selected options on the y-axis) is higher than any other possible combination within the budget constraints.
While it is easy to identify the optimal combination in this toy example, in reality the promotion sets are bigger, and the data arrive in an online manner.

\section{Solution Framework}
\label{sec:solution}

We address the problem with a two-steps approach: an estimation step and an optimization step. The first step is to estimate $\mathit{CATE}_{Y}(i,k)$ and $\mathit{CATE}_{\loss}(i,k)$. The second step is to use the estimations and solve the MCKP and the Online-MCKP.

\subsection{CATE Estimation}

CATE estimation for both $Y$ and $R$ are achieved using uplift modeling and data from randomized controlled trials. We consider different estimators such as two-models~\cite{hansotia2002direct}, transformed outcome~\cite{athey2015machine} and X-learner~\cite{kunzel2019metalearners}. Common evaluation metrics such as Qini Curves and Qini Score are used for model selection.
The step results in estimations of $\mathit{CATE}_{Y}(i,k)$ and $\mathit{CATE}_{\loss}(i,k)$ for every customer $i$ and every promotion $k$. These quantities will serve as the input item sets $K_i$ for the following optimization step.

\subsection{MCKP Approximation Solution}
Similar to the 0-1 knapsack problem, we are interested in an approximation solution to overcome the limitations of computational complexity and lack of fitness for online environments of linear programming solutions. Previous work \cite{zhou2008algorithm} describes an MCKP approximation solution based on Lueker's algorithm \cite{lueker1998average}. We extend the solution and allow for negative values, negative weights, and possible negative budget constraints, which are essential for our business case and problem formulation.
We suggest a four-steps solution that (1) eliminates dominated items; (2) calculates incremental values and weights to allow comparison between items; (3)~transforms the incremental value-weight quantities into efficiency angles to allow efficiency-sorting for both positive and negative values-weights; (4) selects a single item according to an efficiency angle threshold, designed to meet the capacity constraints.

\subsubsection{Dominant items}

Items' dominance plays an important role in solving the MCKP, since it allows to disqualify items that would not be included in an optimal solution. Given an items set $K_i$ we say that an item $b$ is dominated by $a$ if $w_{ia} \leq w_{ib}$ and $v_{ia} > v_{ib}$. Similarly, we say that $b$ is LP-dominated by items $a$ and $c$ if $b$ is dominated by a convex combination of $a$ and $c$. Thus, if $w_{ia} \leq w_{ib} \leq w_{ic}$ and $v_{ia} > v_{ib} > v_{ic}$, then $b$ is LP-dominated by $a$,$c$ if:
$$\frac{v_{ic}-v_{ib}}{w_{ic}-w_{ib}} \geq{\frac{v_{ib}-v_{ia}}{w_{ib}-w_{ia}}}$$
The dominant items of $K_i$ are those that are not dominated or LP-dominated. Dominated items are not expected to be part of an optimal solution since there are items with a higher value and lower weight. 
The approximation solution begins by identifying the set of dominant items  $D_i \subseteq K_i$. The dominant items $D_i$ form the upper-left convex hull of $K_i$ as illustrated in \autoref{fig:toy_efficient}, also known as the Pareto efficiency front. The dominant items can be found in  $\mathcal{O}(|K_i|\cdot log|K_i|)$ time complexity as described in \cite{sinha1979multiple}.

\subsubsection{Incremental value and weight}
Next, for each dominant item $d \in D_i$, we compute its incremental value and weight  $(\overline{v}_{id}, \overline{w}_{id})$. Incremental quantities represent the extra value and extra weight added to the knapsack by choosing item $d$ instead of item $d-1$. This transformation is needed to build a solution that relies on a single efficiency threshold. To compute the incremental quantities we sort the dominant items by increasing weight and proceed as follows:
$$
%\begin{equation}
\overline{w}_{id} = \begin{cases}
w_{id} & if \:\: d=0 \\
w_{id} {-} w_{id-1} & else \\
 \end{cases} \hspace{0.4cm}
\overline{v}_{id} = \begin{cases}
v_{id} & if \:\: d=0 \\
v_{id} {-} v_{id-1} & else\\
 \end{cases}
%  \end{equation}
$$
Adding all the incremental values and weights  $\{(\overline{v}_{id}, \overline{w}_{id})\}$
up to item $d=d^{'}$ to the knapsack is equivalent to adding the original value and weight $(v_{id^{'}},w_{id^{'}})$ of item $d^{'}$.
Another interesting property is that for all $d \geq 1$, the incremental efficiency  $\frac{\overline{v}_{id}}{\overline{w}_{id}}$ is monotonically decreasing with $d$. The property does not hold between $d=0$ and $d=1$ due to negative weights and values, opening the possibility that for $k=0$ the incremental efficiency might be negative. This discontinuity breaks the algorithm proposed by Zhou and Naroditskiy \cite{zhou2008algorithm} and requires us to make a further adaptation in the solution procedure - the efficiency angle.

\begin{algorithm}[b]
	\caption{Efficiency Angle Threshold}
	\label{alg:updatethreshold}
	\begin{algorithmic}[1]
    \State Input:
	    \begin{itemize}
	    \item Set of past dominant items $P$
        \item Knapsack capacity $C$
        \item Current customer index $i$
        \item Expected number of customers $|U|$
        \end{itemize}
    \State Sort $P$ by decreasing angle $\theta_{p}$ 
    \For{$p \in P$}:
    \State \textbf{if} p=0: $f(\theta_{0}) = w_0/|P|$
    \State \textbf{else:} $f(\theta_{p}) = f(\theta_{p-1}) + w_p/|P|$
    \EndFor
    \State \textbf{Return} efficiency threshold $\theta^{*}$:
    %\State \indent $\theta^{*}=f^{-1}(\frac{C}{\frac{|P|}{i}\cdot (|U|-i+1)})$
    \State \indent $\theta^* \gets min_{p \in P} \: \left\{\theta_p \mid f(\theta_p) \leq \frac{C}{\frac{|P|}{i} \cdot (|U|-i+1)} \right\} $
	\end{algorithmic} 
\end{algorithm}

\begin{algorithm}[t]
	\caption{Online MCKP}
	\label{alg:fullonlinealgo}
	\begin{algorithmic}[1]
	    \State Input:
	    \begin{itemize}
	    \item Customer set $U$
        \item  Item sets $K_i$
        \item Knapsack capacity $C$
        \end{itemize}
        \State  $P  \gets \emptyset$
		\For {$(i \in U \mid 1 \leq i \leq |U|)$}
		    \State $D_i \gets$ dominant items of $K_i$ sorted by increasing weight
		    \For {$d \in D_i$}
		    \State Compute incremental values and weights $(\overline{v}_{id},\overline{w}_{id})$:
		    \State \textbf{if} d=0: \indent $\overline{w}_{i0} \gets w_{i0}; \ \overline{v}_{i0} \gets v_{i0}$
            \State \textbf{else}:
		    \indent $\overline{w}_{id} \gets w_{id} {-} w_{id-1};\  \overline{v}_{id} \gets v_{id} {-} v_{id-1}$

		    \State Compute efficiency angle $\theta_{id}$:
		    \State  $
\theta_{id} = 
\begin{cases} \frac{3\pi}{2} & if \:\:\: \overline{v}_{id}=0 \  \land \  \overline{w}_{id}=0\\ 2\pi + atan2(\overline{v}_{id},\overline{w}_{id}) \ &  if  \:\:\: \overline{v}_{id}<0 \  \land \  \overline{w}_{id} \leq 0\\
atan2(\overline{v}_{id},\overline{w}_{id}) & else\\ 
\end{cases}$
		    
		    \State  $P \gets P \cup (\theta_{id}, w_{id})$
	    \EndFor
		  %  \State Sort $P$ by decreasing angle $\theta$ 
		  %  \State Compute threshold function $f$ from items in set $P$:
		  %  \State \indent $f(\theta_{1}) = w_1/|P|$
		  %  \State \indent $f(\theta_{r}) = f(\theta_{r-1}) + w_r/|P|$
		    \State Get updated efficiency threshold $\theta^{*}$:
		    \State \indent $\theta^{*} \gets \mathcal{A}lgorithm\ref{alg:updatethreshold}~(P,C,i,|U|)$
		    \State Find dominant item $d^{*}$:
		    \State \indent $d^{*} \gets \argmin_{d \in D_i} \{ \theta_{id}\: \mid \: \theta_{id}\geq\theta^{*} \} $
		     \State Update capacity:
		     \State\indent $C \gets C - w_{id^{*}}$
		     \State Pick item $d^{*}$
		    
		\EndFor
	\end{algorithmic} 
\end{algorithm}

\subsubsection{Efficiency angle}

Next, we compute the efficiency angle:
%
%\begin{equation}
$$
\theta_{id} = 
\begin{cases} \frac{3\pi}{2} & if \:\:\: \overline{v}_{id}=0 \  \land \  \overline{w}_{id}=0\\ 2\pi + atan2(\overline{v}_{id},\overline{w}_{id}) \ &  if  \:\:\: \overline{v}_{id}<0 \  \land \  \overline{w}_{id} \leq 0\\
atan2(\overline{v}_{id},\overline{w}_{id}) & else\\ 
\end{cases}
$$%\end{equation}

% $$\theta^* \gets max_{p \in P} \: \{\theta_p \mid f(\theta_p) \leq \frac{C}{\frac{|P|}{i} \cdot (|U|-i+1)} \} $$

%
The efficiency angle $\theta_{id}$ is the angle between the incremental quantities $(\overline{v}_{id}, \overline{w}_{id})$ and the positive weight axis as illustrated in the lower part of \autoref{fig:toy_efficient}. The higher the efficiency angle, the smaller the weight of the respective dominant item, a property needed for creating an efficiency angle function as explained in the next section.

\subsubsection{Efficiency angle threshold}
\label{sec:threshold}
The core of the solution method is the efficiency angle function. Its role is to map an efficiency angle $\theta$ to the expected weight of the dominant items with an efficiency angle above $\theta$. The function encodes the distribution of the dominant items from customers we already encountered. We can use that information to decide which promotion to pick for future customers, given our updated budget constraints. The idea behind this function is to allow us to pick an efficiency angle threshold, such that the sum of all the expected weights of future dominant items above the threshold is equal to the remaining knapsack capacity.
We create the efficiency angle function by taking all dominant items from already seen customers into an item set $P$. We sort the items $p \in P$ by decreasing efficiency angle $\theta_p$ and calculate the efficiency angle function as the cumulative average of items' weight:
$$f(\theta_{0}) = w_0/|P|$$
$$f(\theta_{p}) = f(\theta_{p-1}) + w_p/|P|$$
Therefore $f$ is a piece-wise function and can be represented as a list of pairs $\{\theta_{p},f(\theta_p)\}$. The algorithm to create the efficiency angle function is described in Algorithm \ref{alg:updatethreshold}.
%
%\subsubsection{MCKP approximation solution} 
Using the efficiency angle function $f$ we find an approximate MCKP solution. Given a knapsack capacity $C$ and efficiency angle function $f$, we retrieve the efficiency angle threshold $\theta^{*}(i)$ as follows:
%
%$$\theta^{*}(i) =f^{-1}(\frac{ C}{\frac{|P|}{i} \cdot (|U|-i+1)})$$

$$\theta^*(i) = min_{p \in P} \: \left\{\theta_p \mid f(\theta_p) \leq \frac{C}{\frac{|P|}{i} \cdot (|U|-i+1)} \right\} $$
Here, $|U|$ is the total number of customers, $i$ is the current iteration and $|P|/i$ is the average number of dominant items per customer, based on previously observed data. We use the efficiency angle threshold $\theta^{*}$ to find the dominant item on $K_i$ whose efficiency angle is minimal and greater than $\theta^{*}$.
Intuitively, selecting such item will achieve the best value, given the capacity constraints. Picking an item with a smaller angle is not expected to meet the constraint, while picking an item with a higher angle will achieve a sub-optimal value.
A full description of the overall method is presented on Algorithm  \ref{alg:fullonlinealgo}.

\subsection{Algorithm Complexity}

The presented On-MCKP algorithm, without the threshold update step, has an $\mathcal{O}(|K_i|\cdot log|K_i|)$ runtime complexity \textit{per evaluated customer} \cite{sinha1979multiple} and is applicable in an online manner. The threshold update has an $\mathcal{O}(|U|)$ runtime complexity, given that the list of dominant items $P$ is maintained in a sorted data-structure. However, the efficiency angle function update can be executed separately from the real-time calculation, often in a batch-manner.

\begin{figure}[b]
    \includegraphics[width=0.99\columnwidth]{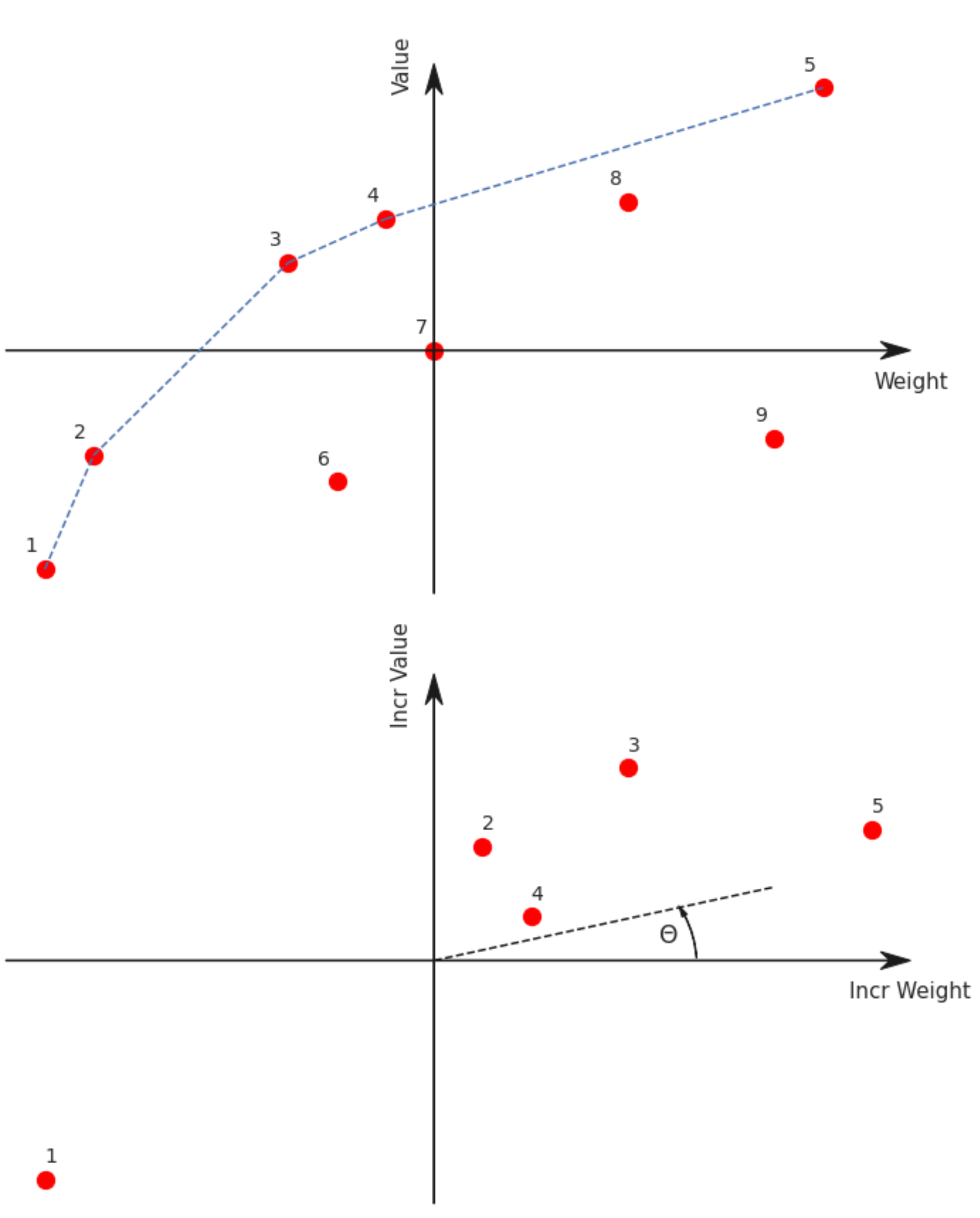}
       \caption{Dominant items selection  (top) and their respective incremental items (bottom). Efficiency angle is in counterclockwise direction from the incremental weight axis.}
        \Description{Dominant items selection example (top) and their incremental items (bottom). Efficiency angle is in counterclockwise direction from the incremental weight axis.}
    \label{fig:toy_efficient}
    \centering
    % source: https://docs.google.com/presentation/d/13MQZgr_69YbTVsWMY4z_dKOQPoFkSVUuvir0UgAXqH4/edit#slide=id.gd600a9cd7e_0_788

\end{figure}

\subsection{Solution Toy Example}

We demonstrate the first steps of the solution method for a single customer on a toy example in  \autoref{fig:toy_efficient}. At the top figure, we can see nine promotion items represented as $(value, weight)$ pairs. Promotions 1 to 5 are the dominant items, forming an upper-left convex hull. We can see that items 6, 7, and 9 are dominated while item 8 is LP-dominated. Dominant items 1 to 5 are sorted by increasing weight. The bottom figure shows the respective incremental values and weights of the dominant items. The efficiency angle $\theta$ starts at the positive weight axis and increases counter-clockwise. We can see that item 1 has the greatest efficiency angle while item 5 has the smallest. This property results from deriving the incremental quantities from dominant items that are sorted by increasing weight. The collection of efficiency angles from different customers is used to compute the efficiency angle function described in section 3.2.4.

\section{Experimental Study}
\label{sec:experiment}

\subsection{Randomized Controlled Trial} 

Prior to performing promotions assignment optimization, we conducted a randomized controlled trial to assess the potential impact of different promotions.  
The experiment took the form of an online multi-variant A/B/N test, in which different treatment groups were offered different promotions while the control group was offered with no promotion.
The experiment was conducted on real traffic of Booking.com website and lasted for several weeks.
In our experiment, the customers in the treatment groups received three different levels of discounts on Booking.com products.
The target metrics - completing a purchase, promotion cost, and incremental revenue - were aggregated and compared between the control and treatment groups, producing an estimation of the average treatment effect on purchase completion and net revenue per treatment.
The experiment showed a conclusively positive treatment effect on purchases for all three promotions but with a conclusive net revenue loss. Such results signal that all the promotion versions require a budget to operate and are not self-sufficient. We are interested in a promotions personalization solution that will allow a zero-budget promotional campaign to operate in the long term.

\subsection{Uplift Modeling on Experimental Data}
\label{sec:upliftresults}

\begin{figure}[t]
    \includegraphics[width=0.95\columnwidth]
    {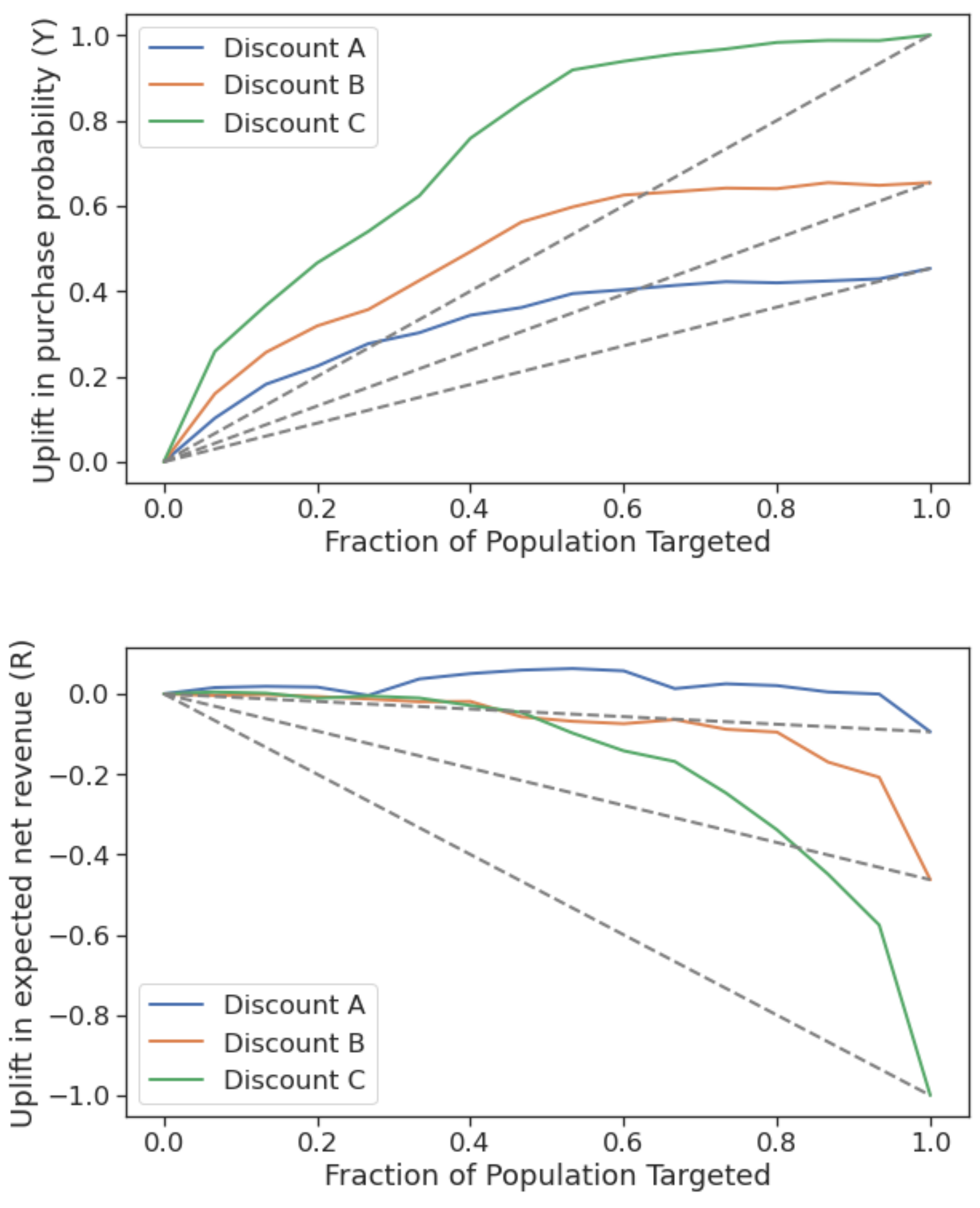}
        \caption{Qini curves of uplift modeling on Y (top) and R (bottom) for 3 discount types (normalized axes)}
       \Description{Qini curves of uplift modeling on Y (top) and R (bottom) for 3 discount types (normalized axes)}
    \label{fig:qini_curves}
    \centering
\end{figure}

The randomized controlled trial mentioned before resulted in a dataset of more than 20 million entries. Each data point was represented by the binary variable $Y$ indicating the completion of a purchase, the continuous variable $R$ indicating the total net revenue (including the promotional costs), and covariates $X$ filled with the customer characteristics.
We selected the best model from a number of uplift modeling techniques in order to obtain $\mathit{CATE}_{Y}(i,k)$ and $\mathit{CATE}_{\loss}(i,k)$. The models were trained on a portion of the dataset, and the predictions were made on another disjoint portion.

The best modeling technique was chosen according to the highest Qini Score on the test set. The normalized Qini plots on the test set for both $Y$ and $R$ are presented on \autoref{fig:qini_curves}. We can see that each discount type has a clear treatment effect when treating 100\% of the population and that the higher the positive effect on $Y$, the higher the negative effect on $R$. We can also see that our models successfully rank the customers by predicted $CATE$ on both $Y$ and $R$. This is evidenced by the shape of the Qini curves and the area below the curve compared to the random assignment line (Qini Score).

\subsection{Datasets}

In addition to the experimental results dataset described before, we tested our proposed method on synthetic and publicly available datasets. The properties of each of the evaluated datasets are described in \autoref{tab:datasets}.

\subsubsection{Experimental data}
Following the uplift modeling procedure described in \autoref{sec:upliftresults}, we generated a sub-sampled dataset of 200,000 sampled customers. For each customer $i$ and each potential promotion $k$ we calculated the expected value ($\mathit{CATE}_{Y}(i,k)$) and the expected weight ($\mathit{CATE}_{\loss}(i,k)$) based on the models predictions. This resulted in $|U| \times K_i$ rows, where the number of treatments per customer $m_i=4 \: \forall i \in U$ (three promotion levels and a no-discount treatment) as described in the first row (Discounts) in \autoref{tab:datasets}. The joint and the marginal distributions of value and weight across three discount levels are depicted on a normalized scale in the bottom part of \autoref{fig:sim_dist}. We observe items in all four quadrants of the chart, with a vast majority in the first quadrant, meaning that we predict that the promotions have a positive value and a positive weight for most customers.

\begin{table}[b]
\caption{Evaluated datasets' properties}
 \label{tab:datasets}
\centering
 \begin{tabular}{|l|c|c|c|}

 \hline
\textbf{Dataset}   & \textbf{\begin{tabular}[c]{@{}c@{}}Number of\\treatments \end{tabular}} & \textbf{\begin{tabular}[c]{@{}c@{}}Number \\ of customers\end{tabular}} & \textbf{Source}   \\
\hline
\textbf{Discounts} & 4  & 200,000 & \begin{tabular}[c]{@{}c@{}}Online experiment\end{tabular}     \\
\textbf{sim5k9}    & 9  & \:\:\:\:5,000 & Simulation  \\
\textbf{sim10k9}   & 9  & \:\:10,000  & Simulation  \\
\textbf{sim20k9}   & 9  & \:\:20,000  & Simulation  \\
\textbf{sim30k9}   & 9  & \:\:30,000  & Simulation  \\
\textbf{sim50k9}   & 9  & \:\:50,000  & Simulation \\
\textbf{sim100k9}  & 9  & 100,000 & Simulation  \\
\textbf{Hillstrom} & 3  & \:\:64,000  & \begin{tabular}[c]{@{}c@{}}Public dataset \cite{radcliffe2008hillstrom}\end{tabular} \\
\hline
\end{tabular}
\end{table}

\subsubsection{Synthetic data}
 
 We generated a synthetic dataset in order to evaluate the scaling of our methods. We simulated a promotional campaign, offering nine levels of discounts ranging between 0 to 40\% in steps of 5\%. 
 The simulation was performed for different population sizes of 5, 10, 20, 30, 50 and 100 thousands customers. For each customer $i$ and each of the nine promotions $k$ we simulated the expected value (conversion uplift: $\mathit{CATE}_{Y}(i,k)$) and expected weight (incremental net revenue loss: $\mathit{CATE}_{\loss}(i,k)$).
 The conversion uplift was randomly sampled from a normal distribution with $\mu = A * D^2 $ and $\sigma^2 = S*D^2 $, where
 $A$ and $S$
 are two global constants fitted by maximum likelihood estimators (MLE) from real data, and $D$ is the respective discount level. Such distribution assumes a positive correlation between the discount level and the conversion uplift, with a diminishing marginal effect.
 
\begin{figure}[t]
\centering
\begin{minipage}{0.4\textwidth}
  \centering
    \includegraphics[width=1\columnwidth]
    {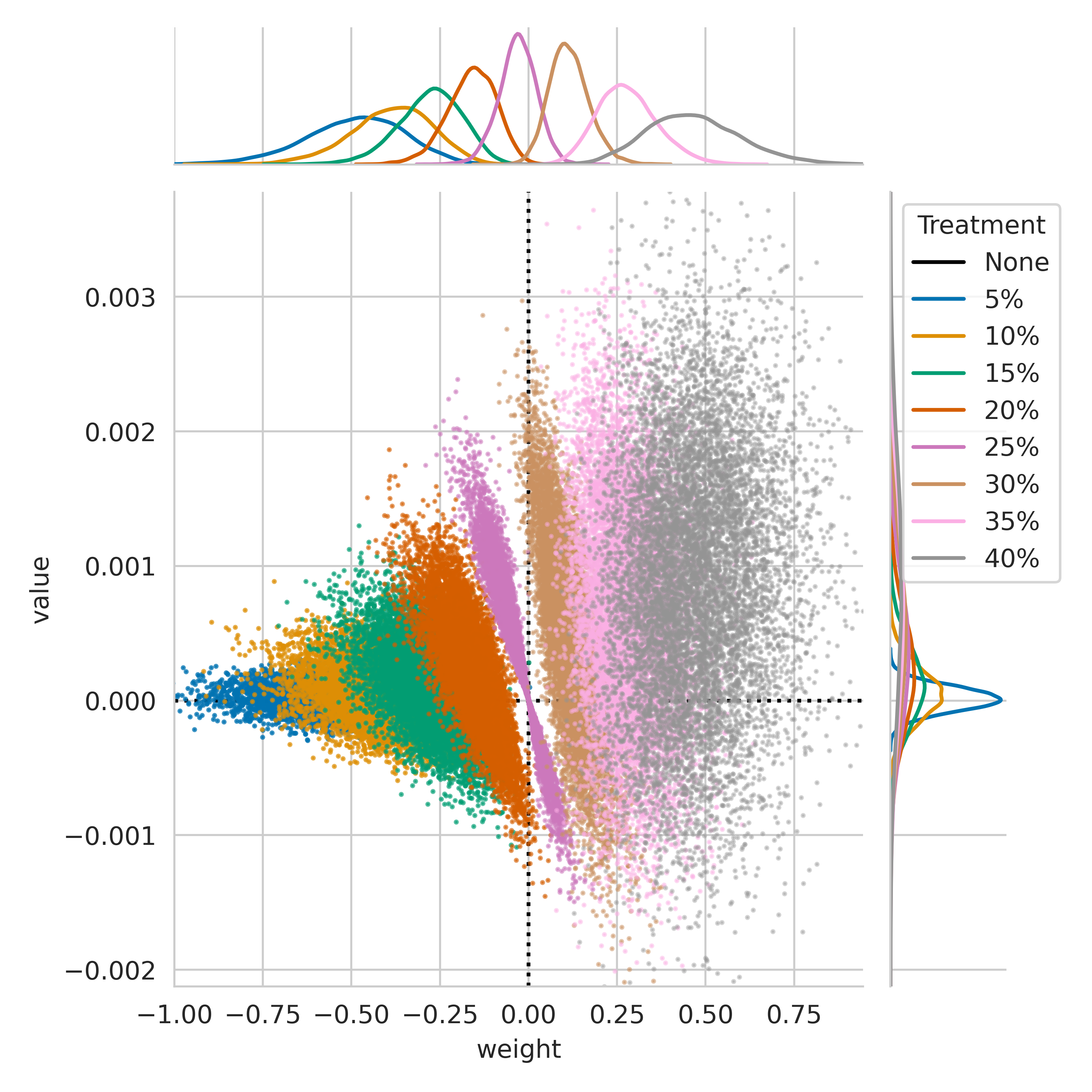}
\end{minipage}
\begin{minipage}{0.4\textwidth}
  \centering
  \includegraphics[width=0.9\textwidth]{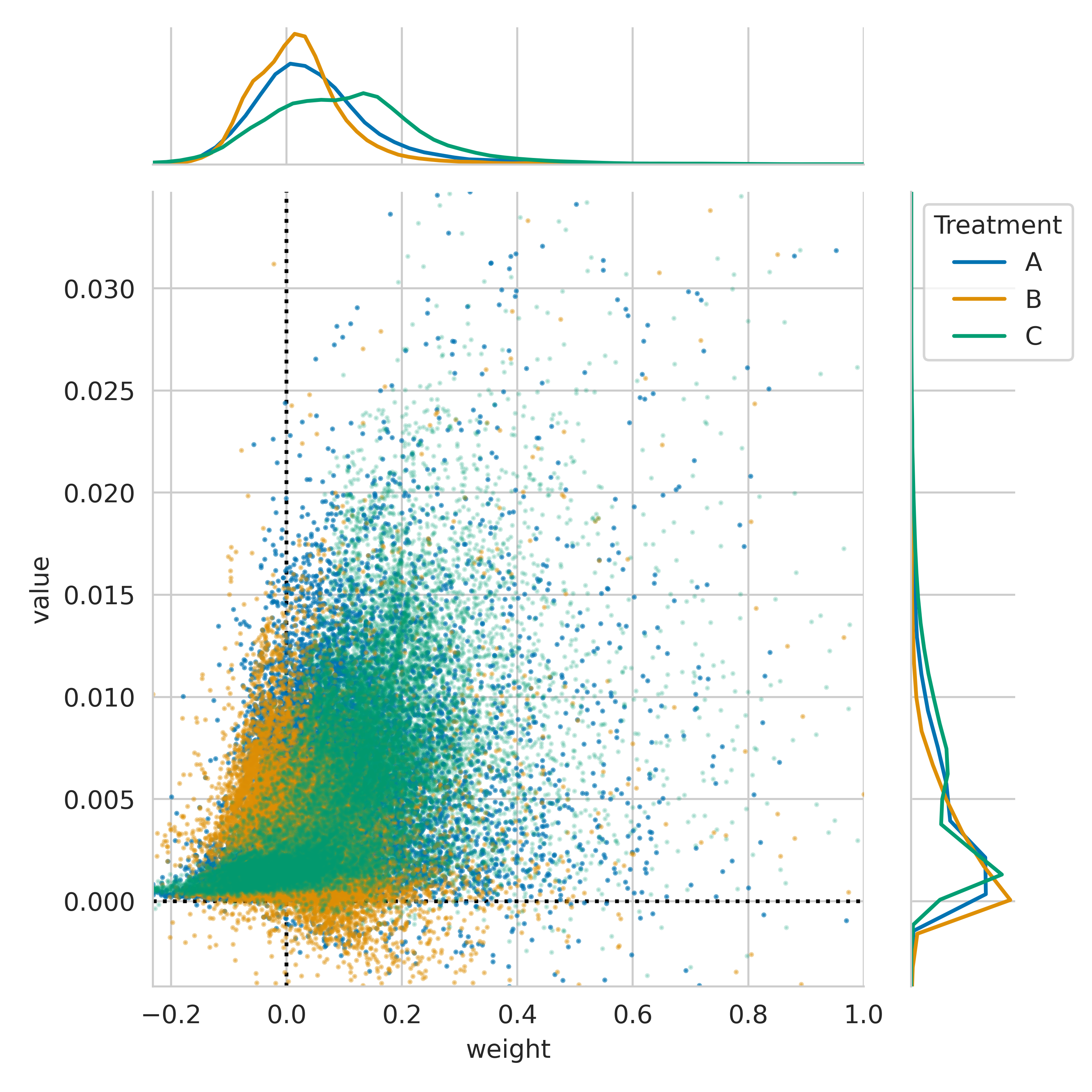}
\end{minipage}
        \caption{Distribution of value and weight of the simulated 10K dataset (top) and real discounts dataset (bottom)}
        \Description{Distribution of value and weights}
    \label{fig:sim_dist}
    \centering
\end{figure}

 The net revenue uplift ($\mathit{CATE}_{R_i}(k,x)= -\mathit{CATE}_{\loss_i}(k,x)$) was sampled from a normal distribution with 
 $ \mu = P*(C-D)$ and $\sigma^2= S_p $ multiplied by $1+\mathit{CATE}_{Y}(i,k)$, estimated by previously sampled conversion uplift. The constants $P$, $C$ and $S_p$ represent MLEs of price, commission and revenue standard deviation respectively.
 The $(C-D)$ multiplier represents the commission loss by applying a discount, and the $[1+\mathit{CATE}_{Y}(i,k)]$ multiplier represents the positive impact of conversion uplift on revenue.
 Therefore, the net revenue loss has a positive correlation with the discount level but can result in negative quantities (profit) due to conversion uplift.
  \autoref{fig:sim_dist} depicts the joint and marginal distributions of value and weight of the simulated promotion dataset for 10,000 customers and nine discount levels (including base variant). Each of the colors represents a different discount level between 5\%-40\%. 
 We notice that higher discounts result in higher average values and weights, introducing a tradeoff between the two dimensions.
 Moreover, we observe a higher variance for both value and weight with the discount level increase, similar to our experimental data.

\begin{figure*}[t]
\centering
\begin{minipage}{0.33\textwidth}
  \centering
  \includegraphics[width=1\textwidth]{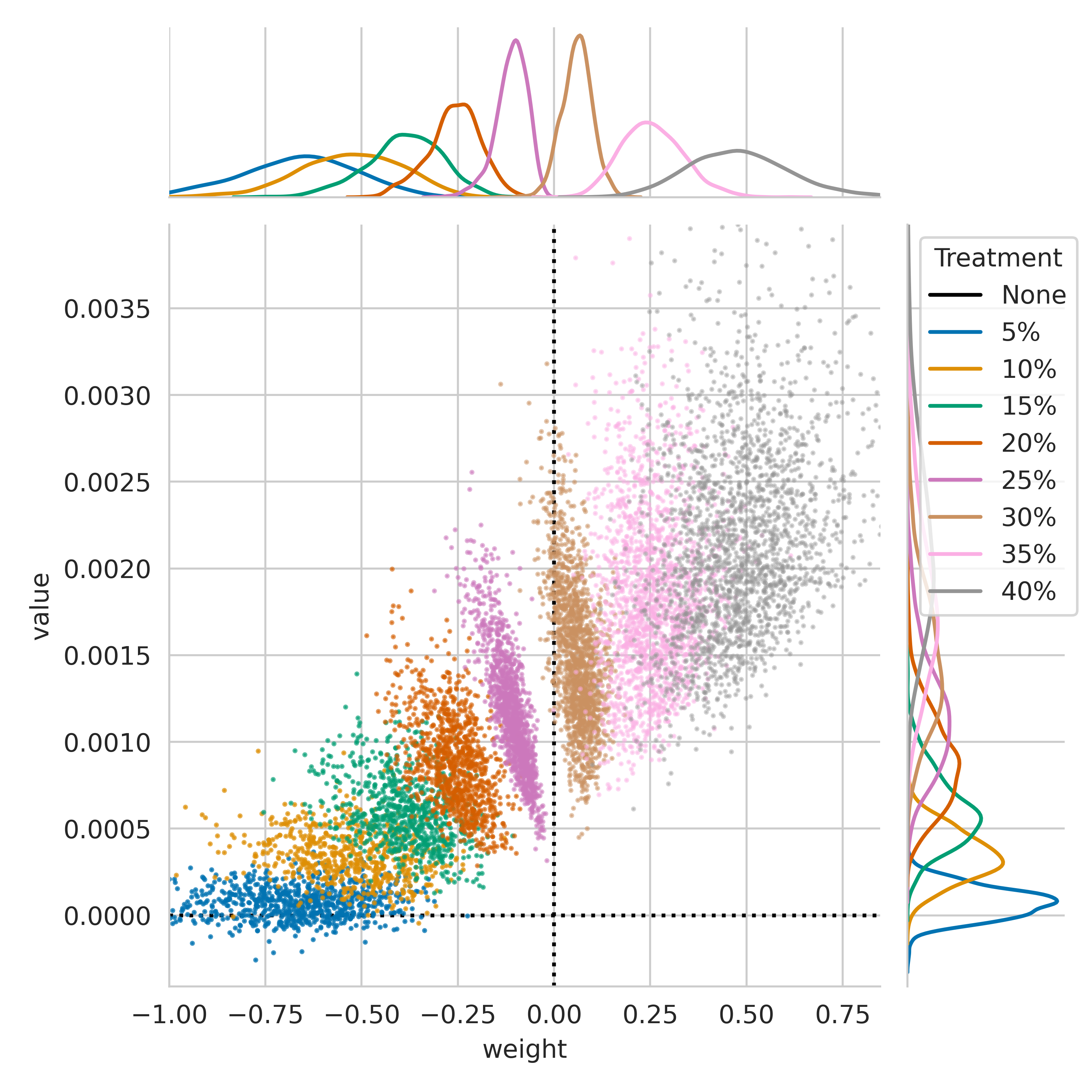}
  \caption*{(a) Simulated 10k dataset}
  \Description{}
\end{minipage}
\begin{minipage}{0.33\textwidth}
  \centering
  \includegraphics[width=1\textwidth]{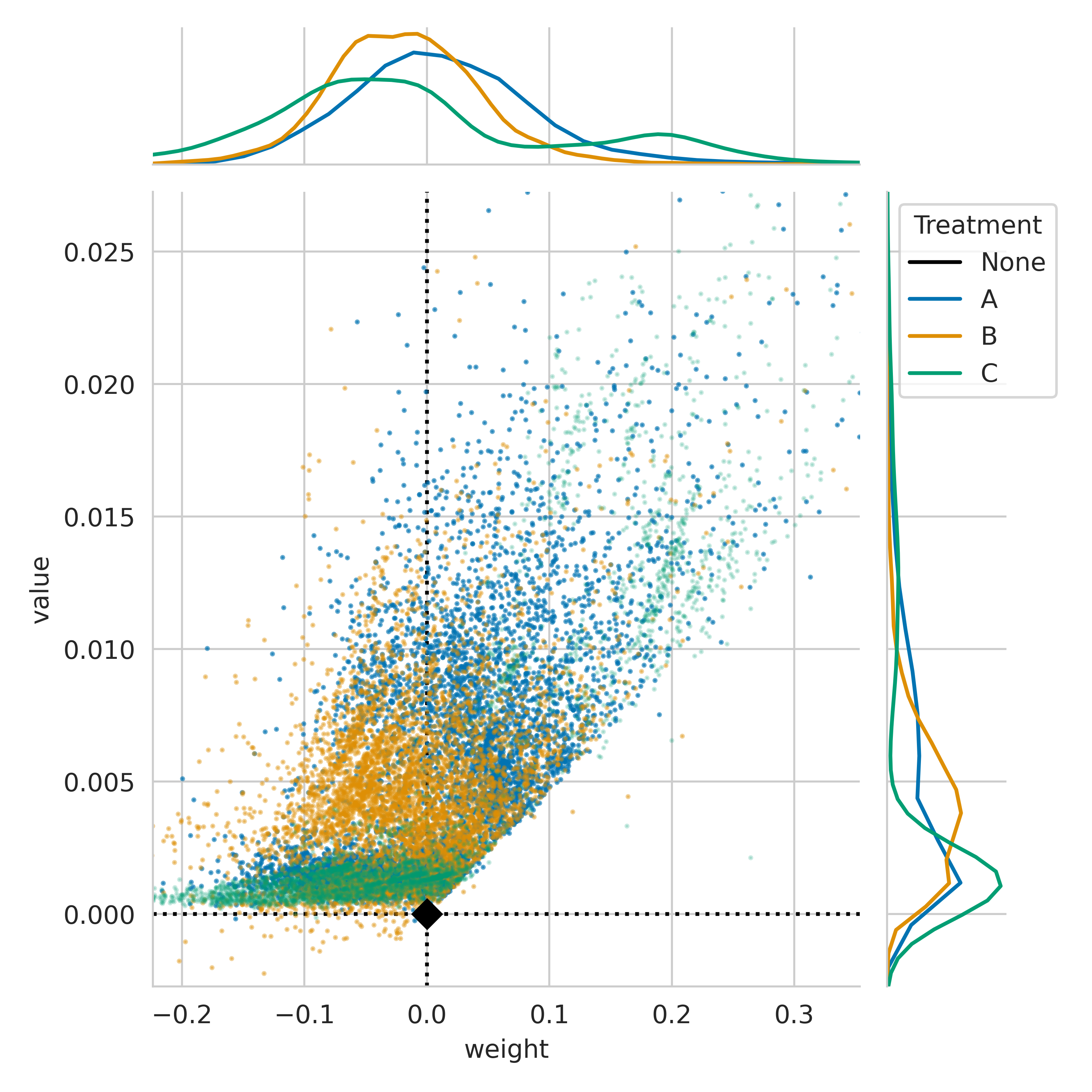}
  \caption*{(b) Real discounts dataset}
  \Description{}
\end{minipage}
\begin{minipage}{0.33\textwidth}
  \centering
  \includegraphics[width=1\textwidth]{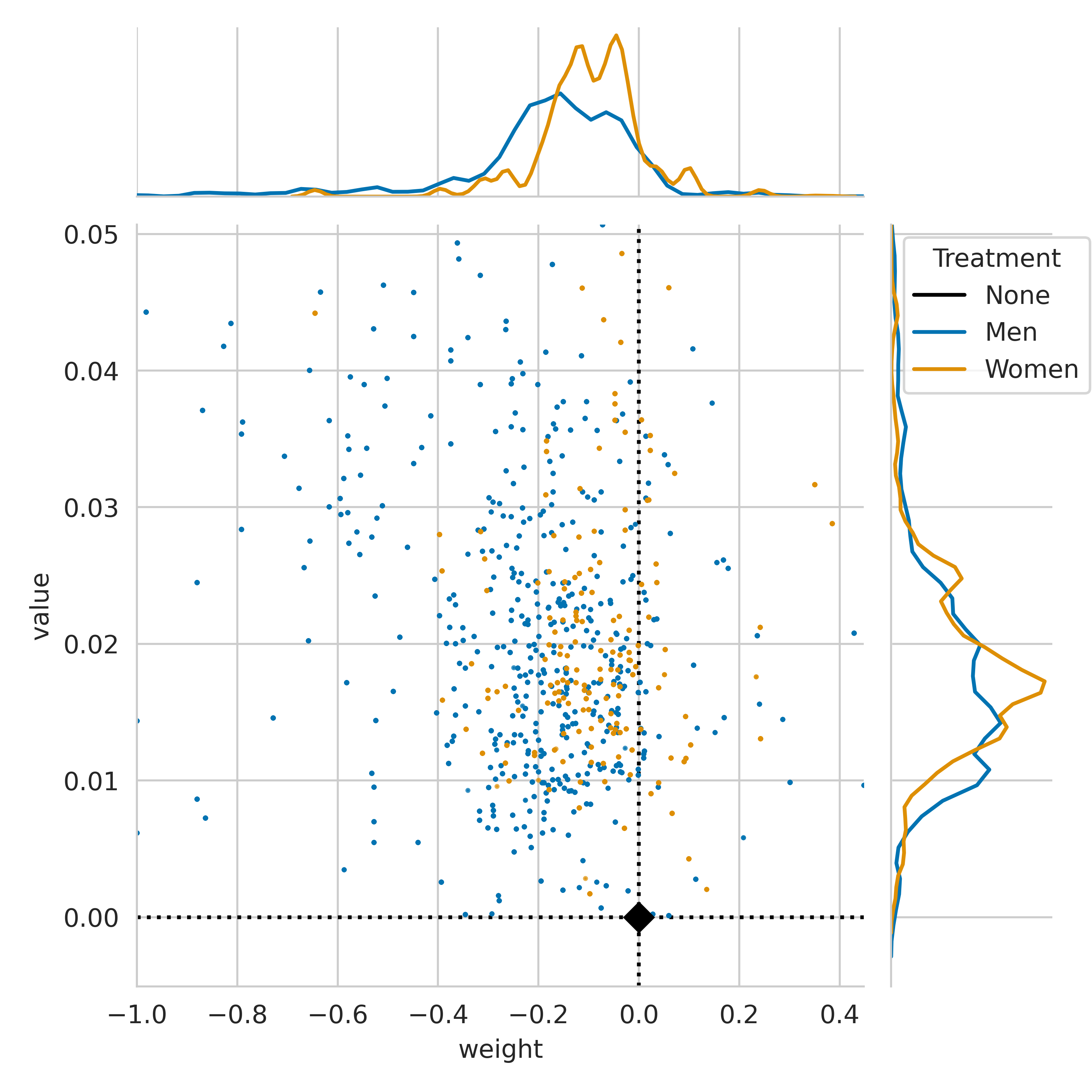}
  \caption*{(c) Hillstrom dataset}
  \Description{}
\end{minipage}
  \caption{Selected promotions in the optimal On-MCKP solution}
  \Description{}
  \label{fig:selection}
\end{figure*}

\subsubsection{Public data}

We evaluated our method on a publicly available e-mail campaign dataset by Hillstrom \cite{radcliffe2008hillstrom} which is commonly used in uplift modeling literature \cite{makhijani2019lore}. The dataset describes an e-mail campaign delivered to 64,000 customers for two weeks.
It consist of an equally random assignment of three treatment variations - \textit{Men merchandise}, \textit{Women merchandise} and \textit{No Treatment}. It contains information about purchase conversion and total spending per customer.
By utilizing the same uplift modeling method as described in \autoref{sec:upliftresults}, we estimated the expected conversion uplift and the expected net revenue loss, which are equivalent to \textit{value} and \textit{weight} in our setup. Similar to our discounts campaign, both conversion uplift and revenue loss could result in negative estimations.
On a side note, while the Hillstrom dataset allows a direct application of our method, there are two fundamental differences in the nature of the data: (1) while discounts imply an associated revenue loss, e-mail campaigns do not have a direct cost factor; (2) Contrary to online web-platforms, e-mail campaigns are an offline marketing channel, and therefore the promotion assignment optimization can be pre-computed in advance.

\subsection{Assignment Optimization Methods}

Our study compares several online and offline optimization methods to estimate the effectiveness of the suggested Online-MCKP solution. The methods were compared on all input datasets, with a zero-budget constraint, namely - the promotion campaigns are required to be self-sponsored.
The compared methods are:

% \begin{itemize}
%     \item Global Selection (\textit{Global})
%     \item Local Selection (\textit{Local})
%     \item Greedy Selection (\textit{Local})
%     \item Online Multi-Choice Knapsack (\textit{On-MCKP})
%     \item Offline Multi-Choice Knapsack (\textit{Off-MCKP})
%     \item Integer Linear Programming (\textit{ILP})
% \end{itemize}
%
% All reported calculations were performed on a single Linux machine, with 8-cores and 10 GB RAM, using Python 3.8 software. Sequential calculations of online algorithms (such as On-MCKP) were performed on a single core.

%\subsubsection{Global Selection}
\subsubsection{Global Selection (Global)}

This method is used as a standard benchmark solution for selecting the best treatment - a classical business strategy via an A/B test, assuming a near-homogeneous treatment effect. The global solution picks a single treatment, out of the possible promotion options, and suggests it to all of the customers in a non-personalized manner. The selected treatment is the one that achieves the highest total value while complying the budget constraint. In cases where none of the suggested treatments meets the budget constraint, the base variant will be the selected solution, without showing a promotion to any of the customers.

%\subsubsection{Local Greedy Solution}
\subsubsection{Local Solution (Local)}

The local solution solves a local optimization problem within the scope of a single customer - it picks the promotion with the highest value and non-positive weight. An example of such a solution is depicted in the toy example in \autoref{fig:toy_example}b. This basic solution relies on the individual impact estimation of each promotion per customer and can be applied online without sharing knowledge between the individual decisions.

%\subsubsection{Local Greedy Solution}
\subsubsection{Greedy Solution (Greedy)}

Similar to \textit{Local}, the greedy solution solves the optimization problem within the scope of a single customer. However, it picks the promotion with the highest value and a weight within the \textit{current capacity state}. This allows to pick items with positive-weight in cases where previously picked items increased the remaining capacity.

\subsubsection{Online MCKP (On-MCKP)}
%\subsubsection{Online MCKP}
This method corresponds to the real-world scenario where customers arrive one at a time, and the decision of which promotion to offer must be made at each time step. At the beginning of the process, we have no information about the general weights and values distributions. The method adapts the promotion assignment decision to the remaining budget and the updated efficiency angle function as described in Algorithm \ref{alg:fullonlinealgo}.
%It is worth noticing that the algorithm can be updated in batches in actual online setups, independently of real-time serving platforms.

%\subsubsection{Offline MCKP}
\subsubsection{Offline MCKP (Off-MCKP)} 
This method uses the same underlying selection algorithm as the Online MCKP, while the items are provided in advance, allowing to fit the efficiency angle function once, prior to the assignment decision.
Then the algorithm decides which promotion to offer to each customer without updating the efficiency angle function. The core of the method, the efficiency angle function, is described in Algorithm  \ref{alg:updatethreshold}.

\subsubsection{Integer Linear Programming (ILP)}
%\subsubsection{Integer Linear Programming}
We use an ILP solver in order to find an optimal upper bound solution in an offline setup. The MCKP can be solved as an ILP using the linear formulation described in \autoref{eq:mckp}. The budget constant $C$ was set to zero ($C=0$).
We relied on python PuLP package \cite{mitchell2011pulp} with CBC (Coin-or branch and cut) solver. The overall solver runtime was limited to 3-hours. On a side note, while there are more suitable solvers for knapsack problems, we only faced a single instance where the optimal solution was not reached within the limited runtime, achieving a feasible solution with an infinitesimal optimality gap.

% \begin{table}[b]
%  \caption{Optimality rate of solutions (compared to optimal ILP) of evaluated methods across different datasets}
%  \label{tab:results}
% \begin{tabular}{lrrrr|}
% \cline{2-5}
% \multicolumn{1}{c}{\textbf{}} &  \multicolumn{1}{|l}{\textbf{Global}} & \textbf{Local} & \textbf{On-MCKP} & \textbf{Off-MCKP} \\ \hline
% \cline{1-1}
%  \multicolumn{1}{|l}{\textbf{Discounts}} & 0.0\% & 37.0\% & 99.75\% & \textgreater{}99.99\% \\ \hline
%  \multicolumn{1}{|l}{\textbf{sim5k9}} & 16.9\% & 50.9\% & 99.99\% & \textgreater{}99.99\% \\ \hline
%  \multicolumn{1}{|l}{\textbf{sim10k9}} & 25.6\% & 50.3\% & 99.98\% & 99.99\% \\ \hline
%  \multicolumn{1}{|l}{\textbf{sim20k9}} & 18.6\% & 46.6\% & \textgreater{}99.99\% & \textgreater{}99.99\% \\ \hline
%  \multicolumn{1}{|l}{\textbf{sim30k9}} & 26.0\% & 50.6\% & 99.99\% & \textgreater{}99.99\% \\ \hline
%  \multicolumn{1}{|l}{\textbf{sim50k9}} & 26.4\% & 51.0\% & 99.99\% & \textgreater{}99.99\% \\ \hline
%  \multicolumn{1}{|l}{\textbf{sim100k9}} & 25.9\% & 50.6\% & 99.99\% & \textgreater{}99.99\% \\ \hline
%  \multicolumn{1}{|l}{\textbf{Hillstrom}} & 91.9\% & 93.3\% & \textgreater{}99.99\% & \textgreater{}99.99\% \\ \hline
% \end{tabular}
% \end{table}

\begin{table}[b]
 \caption{Optimality rate of solutions (compared to optimal ILP) of evaluated methods across different datasets}
 \label{tab:results}
\begin{tabular}{lrrrrr|}
\cline{2-6}
\multicolumn{1}{c}{\textbf{}} &  \multicolumn{1}{|l}{\textbf{Global}} & \textbf{Local} & \textbf{Greedy} & \textbf{Online} & \textbf{Offline} \\ \hline
\cline{1-1}
 \multicolumn{1}{|l}{\textbf{Discounts}} & 0.0\% & 37.0\% & 65.79\% & 99.75\% & \textgreater{}99.99\% \\ \hline
 \multicolumn{1}{|l}{\textbf{sim5k9}} & 16.9\% & 50.9\% & 79.09\% & 99.99\% & \textgreater{}99.99\% \\ \hline
 \multicolumn{1}{|l}{\textbf{sim10k9}} & 25.6\% & 50.3\% & 80.42\% & 99.98\% & 99.99\% \\ \hline
 \multicolumn{1}{|l}{\textbf{sim20k9}} & 18.6\% & 46.6\% & 76.09\% & \textgreater{}99.99\% & \textgreater{}99.99\% \\ \hline
 \multicolumn{1}{|l}{\textbf{sim30k9}} & 26.0\% & 50.6\% & 80.59\% & 99.99\% & \textgreater{}99.99\% \\ \hline
 \multicolumn{1}{|l}{\textbf{sim50k9}} & 26.4\% & 51.0\% & 79.94\% & 99.99\% & \textgreater{}99.99\% \\ \hline
 \multicolumn{1}{|l}{\textbf{sim100k9}} & 25.9\% & 50.6\% & 80.42\% & 99.99\% & \textgreater{}99.99\% \\ \hline
 \multicolumn{1}{|l}{\textbf{Hillstrom}} & 91.9\% & 93.3\% & \textgreater{}99.99\% & \textgreater{}99.99\% & \textgreater{}99.99\% \\ \hline
\end{tabular}
\end{table}

\section{Results}
\label{sec:results}

\begin{figure}[b]
        \includegraphics[width=0.99\columnwidth]{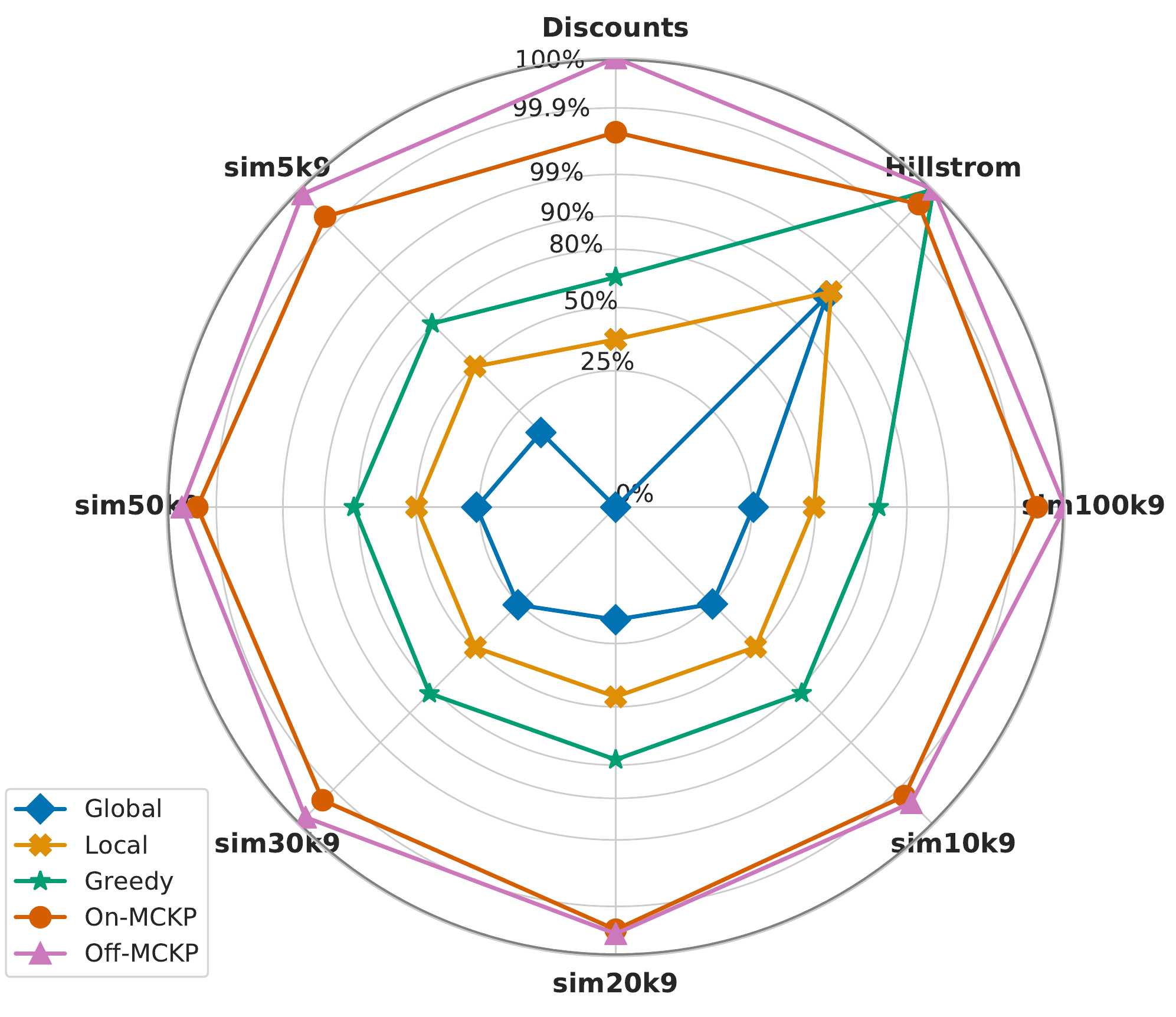}
        \caption{Achieved percentage of the optimal ILP solution across different methods and datasets. (Radial log scale)}
        \Description{Achieved total value of compared methods as a percentage of the optimal solution across all datasets.}
    \label{fig:opt_comp}
\end{figure}

\begin{figure*}[t]
    \includegraphics[width=0.95\textwidth]{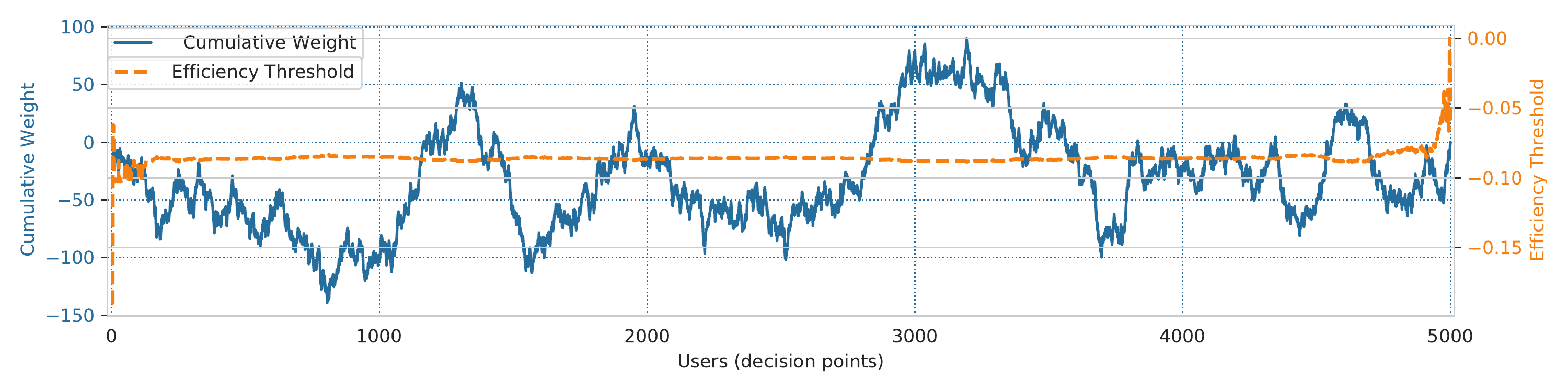}
        \caption{On-MCKP dynamics of total  weight (net revenue loss) and decision threshold on simulated dataset with 5K customers.}
        \Description{}
    \label{fig:dynamics}
    \centering
\end{figure*}

\subsection{Optimization results}

The complete optimization results of the \textit{Global}, \textit{Local}, \textit{Greedy}, \textit{Online} and \textit{Offline MCKP} methods are listed in \autoref{tab:results} and depicted in \autoref{fig:opt_comp}. We report the optimality rate of the total value compared to the optimal ILP solution, across the datasets.

The \textit{Off-MCKP} demonstrates a near-optimal performance with a negligible optimality gap up to $\num{7.4E-05}$.
The \textit{On-MCKP} demonstrates excellent performance as well, with a maximal optimality gap of 0.245\% on the real discounts dataset. We observe that the \textit{Global}, non-personalized benchmark achieves low performance (16-26\%) at the simulated instances and can not find any feasible solution for the real discounts dataset.
The \textit{Local} solution constantly outperforms the \textit{Global}, but is limited compared to the suggested MCKP methods, delivering about 51\% of the expected impact.
The \textit{Greedy} method results in 65\%-80\% optimality rates, and a near-optimal solution in Hillstorm problem, in which the capacity constraint is non-binding.

The suggested MCKP solutions widely outperform the \textit{Global}, \textit{Local} and \textit{Greedy} benchmarks. In the real discounts case, it plays a game-changing role in allowing the promotional campaign to become self-sponsored, with a significant improvement compared to  \textit{Local} (+170\%) and \textit{Greedy} (+51\%) solutions.

\subsection{Solution Properties}

To better understand the properties of our proposed method, we dive into the Online-MCKP solution on three datasets - Simulated 10K customers, Discounts and Hillstrom as depicted in \autoref{fig:selection}.
Following the earlier visualization of the available promotions of \textit{sim10k9} dataset in \autoref{fig:sim_dist}, we present the optimal composition of promotion assignments on the same dataset in \autoref{fig:selection}a.
We observe an interesting phenomenon where the selected promotions follow a linear value/weight trend, forming a separation bound by excluding the solutions below the line which is equivalent to the efficiency threshold used by the method.
Similarly, the real discounts dataset in \autoref{fig:selection}b forms a separation bound and depicting a clear tradeoff between value and weight. It consists of a blend of all three available promotions, using items from three possible quadrants.
The Hillstrom dataset in \autoref{fig:selection}c  demonstrates a scattered spread of the selected promotions. This is because the email campaign does not imply a direct revenue loss, and therefore, the capacity constraint is not bounding this solution. The optimal solution does not contain any promotion with a negative value since it always preferred to pick the highest-value option per customer.

It is worth noticing that all the solutions do not include items from the fourth (negative-value/ positive-weight) quadrant, since all these items are dominated by the default (0,0) option.  The discounts and Hillstrom solution compositions include selecting the base (no-promotion) treatment (1\% in Hillstrom and 19.7\% in Discounts). In the simulated dataset we never preferred to pick the default no-promotion item over the other eight alternatives.

\subsection{Online Dynamics}

We demonstrate the online dynamics of the On-MCKP algorithm on the simulated 5K customers dataset. \autoref{fig:dynamics} depicts the step-by-step updates of the efficiency threshold (orange scale) and the cumulative weight (blue scale) through the 5,000 iterations.
The cumulative weight fluctuates around the target capacity, exceeding the capacity constraint at certain time-steps, but quickly recovers to the zero-target. 

The efficiency angle threshold presents high fluctuations in the first steps, and quickly converges to a stable efficiency angle around $\theta=-0.085$.
This angle relates to the observed separation boundary observed in \autoref{fig:selection}a.
Interestingly, the threshold changes drastically in the final rounds. Since the threshold is highly dependant on the expected weight of the remaining items, its efficiency angle returns to near-zero levels once the algorithm does not expect any additional users.
In real use-cases, we are not supposed to observe such behavior since the self-sponsored promotional campaign is expected to run for an unlimited time horizon.

\section{Conclusion}
\label{sec:conclusion}

Our study presents a novel formulation of \emph{Online Constrained Multiple-Choice Promotions Personalization Problem}, relying on the Multiple-Choice Knapsack Problem.
It suggests a two-steps framework based on causal uplift estimation and online constrained optimization.
We extend the literature on the multiple-choice knapsack problem by providing a solution for negative values and weights, which are common in causal estimations and especially useful with the zero-budget constraint.
Our experimental study validates that the suggested method is a highly effective solution to the problem and applicable for decision mechanisms in real-time e-commerce setups.
The method provides an adaptive solution for a dynamic environment and achieves above 99.7\% of the possible optimal promotional impact. It ensures budget constraints compliance by responding to budget offsets caused by past decisions.
The personalized assignment of promotions is a game-changer for scaling self-sponsored marketing campaigns, as it increases the promotion's impact by orders of magnitude. In our promotional business case, the use of personalization transformed a campaign with an insufficient budget into an effective, long-term and self-sponsored application.

The presented optimization framework can be further generalized to allow for different number and types of promotions per customer and does not require a special domain knowledge about the nature of the offer. 
Our angle-transformation solution for negative values opens up potential research directions for other knapsack algorithms. Describing the items in the two-dimensional, value-weight manner allows extending the applications of the method beyond the e-commerce industry, such as in industrial production, traffic solutions, trading algorithms, and more.
We expect further potential improvements in the method by converting the estimation-optimization framework into an end-to-end tailor-made learner.
We look forward to validating the holistic framework by conducting an online comparison study and encourage the research community to test the method on additional use-cases and extend the research on budget-constrained personalization problems.

% \begin{acks}
% value team
% research group
% internal reviewers

% \end{acks}

\balance
\bibliographystyle{ACM-Reference-Format}
\bibliography{bibliography}

\end{document}